\documentclass{article}
\usepackage{spconf,amsmath,graphicx}
\usepackage{algorithm}
\usepackage{algpseudocode}
\usepackage{multicol}
\usepackage{hyperref}
\usepackage{xcolor}
\usepackage{float}
\usepackage{multirow}
\usepackage{optidef}
\usepackage{cases}
\usepackage[mode=build]{standalone}

\usepackage[letterpaper, left=0.75in, right=0.60in, bottom=0.60in, top=1in]{geometry}

\title{Multiple description video coding for real-time applications using HEVC}%
\twoauthors
 {Trung Hieu Le, Marc Antonini, member IEEE \thanks{This project is funded by the Région Sud, France and LEXTAN SAS.}}
	{I3S laboratory, Côte d’Azur University and CNRS\\ 
	UMR 7271, Sophia Antipolis, France}
 {Marc Lambert, Karima Alioua}
	{LEXTAN SAS\\
	Gemenos, France}
\begin{document}
%
\maketitle
\begin{abstract}
\vspace{-0.4\baselineskip}
Remote control vehicles require the transmission of large amounts of data, and video is one of the most important sources for the driver. To ensure reliable video transmission, the encoded video stream is transmitted simultaneously over multiple channels. However, this solution incurs a high transmission cost due to the wireless channel's unreliable and random bit loss characteristics. To address this issue, it is necessary to use more efficient video encoding methods that can make the video stream robust to noise. In this paper, we propose a low-complexity, low-latency 2-channel Multiple Description Coding (MDC) solution with an adaptive Instantaneous Decoder Refresh (IDR) frame period, which is compatible with the HEVC standard. This method shows better resistance to high packet loss rates with lower complexity.
\end{abstract}
\vspace{-1.0\baselineskip}

\section{Introduction}
\vspace{-1.0\baselineskip}
Remote driving requires continuous video data transmission that allows the driver to perceive the environment. The video from the vehicle is transmitted to the driver through a wireless channel. In such an application, the latency and quality of the frame must respect some limit to guarantee safety. In \cite{DBLP:journals/comsur/ParvezRGSD18} it is shown that the latency requirement may vary between 10ms to 100ms and depends on the speed of the vehicle. The emergence of 5G networks, with their high-performance and low latency capabilities, makes it possible to transmit high quality video in real-time for remote driving applications. However, compared to data transmission through wired networks, wireless networks are more subject to noise and therefore remain less reliable.

Several techniques for Forward Error Correction (FEC) have been implemented in the transport layer to improve the resilience of video transmission to losses. These techniques, such as Low Density Parity Check (LDPC) or Turbo code (TC), involve adding redundant correction data to the bitstream in order to correct corrupted packets. These techniques make it possible to reach the Shannon limit. However, FEC is designed to be robust only up to a certain limit of packet error rate, and requires high complexity computation \cite{1698467}. Moreover, in a mobility context, the wireless channel's characteristics are dynamic and vary over time, making it difficult to estimate the channel state accurately for FEC.

In the context of real-time video transmission, one possible approach to mitigate interference is to use two separate wireless channels to send the same video sequence as a backup. However, this approach can result in a waste of bandwidth since the same information is being transmitted twice. To address this issue, Multiple Description Coding (MDC) can be used as a solution. MDC was first introduced in 1979 by Gersho, Witsenhausen, Wolf, Wyner, Ziv, and Orarow and appears to be a promising solution to this problem \cite{DBLP:journals/tit/GamalC82a}. It was shown to be an effective error-resilient under packet error, random bit error and routing delay in wireless condition \cite{DBLP:journals/spm/Goyal01a}. 

MDC of an image involves encoding multiple representations or descriptions of the image such that if some of the descriptions are lost or corrupted, the remaining descriptions can still be used to reconstruct the original image with some degree of quality degradation. Specifically in the case of MDC for two channels, the MDC encoder produces two different descriptions, S1 and S2, of the video with bit rates R1 and R2, respectively, from the original video source. These two descriptions are then transmitted through two independent channels by two transmitters. If only one description is available at the MDC decoder, either S1 or S2, the side decoder will be used to produce the video sequence with distortion D1 or D2, respectively. However, if both descriptions are available at the MDC decoder, the central decoder merges them to construct the so-called central reconstruction by removing redundant information and retaining the primary one. As a result, the video quality will be higher with a smaller central distortion D0.

MDC can be classified into three categories based on how the redundancy information is added to each description \cite{DBLP:journals/mms/KazemiSS14}: MDC in the spatial domain, MDC in the frequency domain, and MDC in the time domain. There have been many studies on MDC in the spatial domain, most of which are compatible with the H.264 encoding standard. In \cite{DBLP:journals/tcsv/TilloGO08}, Tilo {\it et} al proposed a method of MDC compatible with the H.264 standard based on adjusting the redundancy level of the different slices.
In \cite{DBLP:journals/tcsv/RadulovicFWHH10}, the author used the Multiple-State Video Coding with Redundant Pictures method by shifting the I and P frames of Group of Picture (GOP) between the two descriptions. In \cite{DBLP:journals/tcsv/HsiaoT10}, the authors proposed an H.264 MDC scheme based on block permutation and DCT coefficients splitting. 

On the one hand, the H.264 standard is widely used for video coding. On the other hand, the HEVC (High Efficiency Video Coding) standard has higher performance, being able to reduce the bit rate by up to 50\% compared to H.264 and more resilience to the packet erasure, as demonstrated in \cite{DBLP:journals/jrtip/Psannis16}. This is achieved through the use of a coding structure called the Coding Tree Unit (CTU), which can be divided into smaller Coding Units (CU) based on the desired bitrate and quality target \cite{DBLP:series/icas/SchwarzSM14}. Many recent studies have attempted to include a multiple description coding (MDC) scheme in the HEVC standard, such as in \cite{DBLP:journals/mta/MajidOA18,DBLP:journals/jvcir/WangCZC22} which are based on the visual saliency and transform domain splitting. However, these methods are not suitable for real-time applications due to their use of Bidirectional Motion Compensation (B-Frame). Authors in \cite{LABIOD201923} proposes a frame rate variation MDC scheme adapted to the remote control vehicule. Nevertheless, this scheme didn't exploit the temporal dependency between frames, thus the compression performance is low.

In this article, we propose a spatial domain multiple description codec based on the HEVC standard with Rate Distortion Optimization (RDO) at the CTU level for each frame independently. Our scheme uses unidirectional temporal prediction (T0 layer motion compensation), which helps reducing the transmission cost and meet real-time delay requirements. This paper is divided into two parts. The first part presents the proposed solution, and the second part shows its performance under packet erasure channel.
\vspace{-1.0\baselineskip}

\section{Proposed Method}
\vspace{-0.8\baselineskip}
\subsection{MDC decoding workflow}
\begin{figure}[!ht]
\centering
\includegraphics[width=0.50\textwidth]{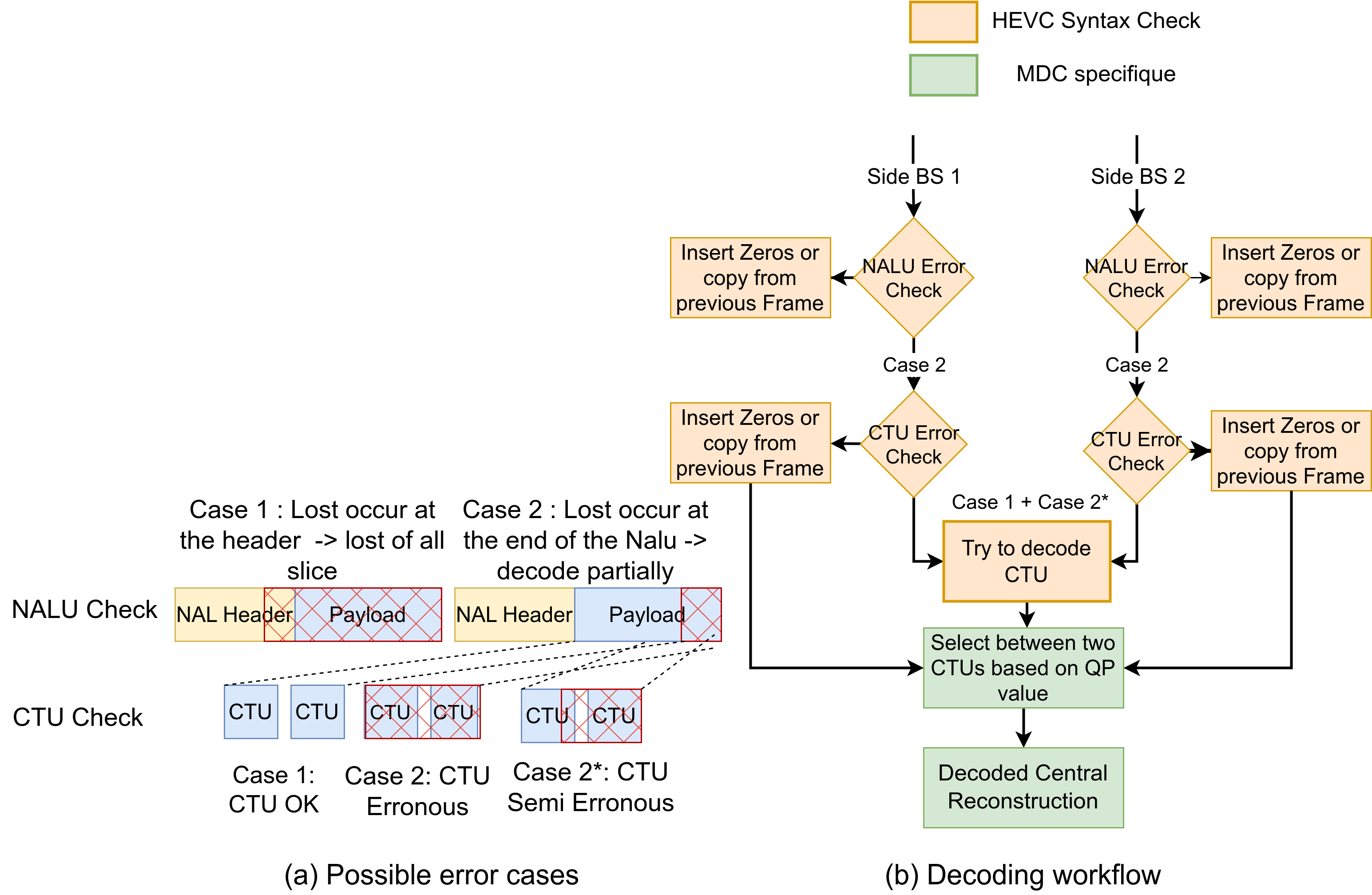}
\vspace{-1.5\baselineskip}
\caption{\footnotesize{MDC Decoder}} 
\label{fig:DecodeWorkFlow}
\vspace{-1.5\baselineskip}
\end{figure}

\vspace{-0.5\baselineskip}
In order to detect the error in the bitstream, we build a decoder with a syntax checker as proposed in  \cite{DBLP:conf/mmsp/SabevaJKD06}. Based on the syntax, and the a priori information from the header, we can distinguish the different of error cases. Then, we propose the following MDC decoding workflow (see figure \ref{fig:DecodeWorkFlow}):
\vspace{-0.5\baselineskip}
\begin{enumerate}
\item Create two zero buffers which will contain two side pictures.
\vspace{-0.5\baselineskip}
\item Check at the Network Abstraction Layer Unit (NALU) level of each description. If an error is detected in the header of the NALU, then discard the NALU.
\vspace{-0.5\baselineskip}
\item If no error is detected at the NALU header, the decoder continues to decode each CTU of each NALU, and an error check is performed at the CTU level. If an error is detected in a CTU during the decoding process, then discard all the following CTUs (after the erroneous one) which belong to this NALU.
\vspace{-0.5\baselineskip}
\item At the end, when the complete frame is received, discard all the zeros and replace them with non-zeros. If two CTUs of two descriptions are erroneous at the same time, then replace with the CTU in the previous frame.
\vspace{-1.0\baselineskip}
\end{enumerate}

\vspace{-0.5\baselineskip}
\begin{figure*}
    \centering
    \includegraphics[width=0.80\textwidth]{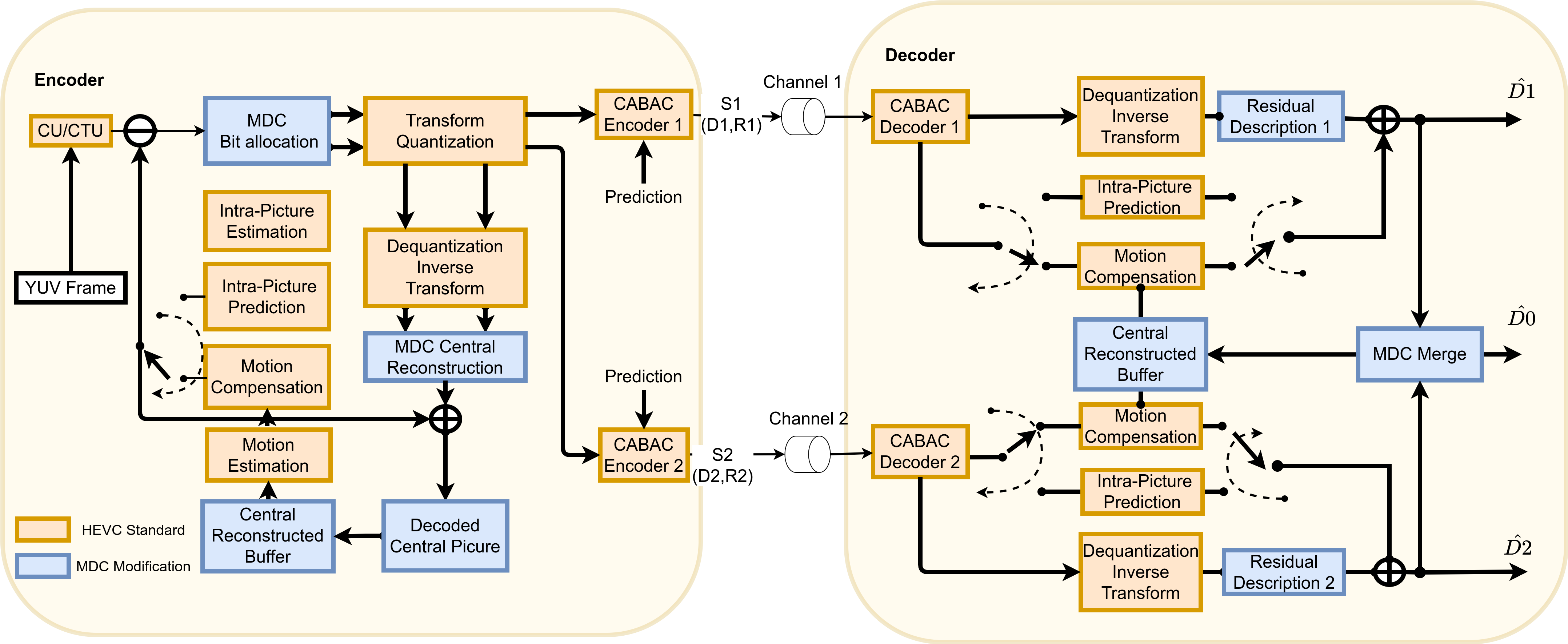}
    \vspace{-1.0\baselineskip}
    \caption{\footnotesize{Proposed coding scheme: First, the frame will be split in CTU/CU structure optimized for the given target bitrate. Based on this structure and the residual produced by this step. The MDC bit allocation will allocate the $QP$ value for each CTU using algorithm \ref{algo:algoSolveLBGS}. Then, these  $QP$ values will feed the Transform and Quantization blocks at encoder side. After that the central reconstruction will be constructed by discarding the coarse quality CTU, this central reconstruction is stored in the Central Reconstructed Buffer. Finally MC process is performed on the central reconstruction.}}
    \vspace{-0.5\baselineskip}
    \label{fig:MyMDCSolution}
\end{figure*}
\subsection{MDC bit allocation algorithm} 
\begin{figure}
    \centering
        \begin{minipage}[b]{0.5\linewidth}
		\centerline{
            \includegraphics[width=2in]{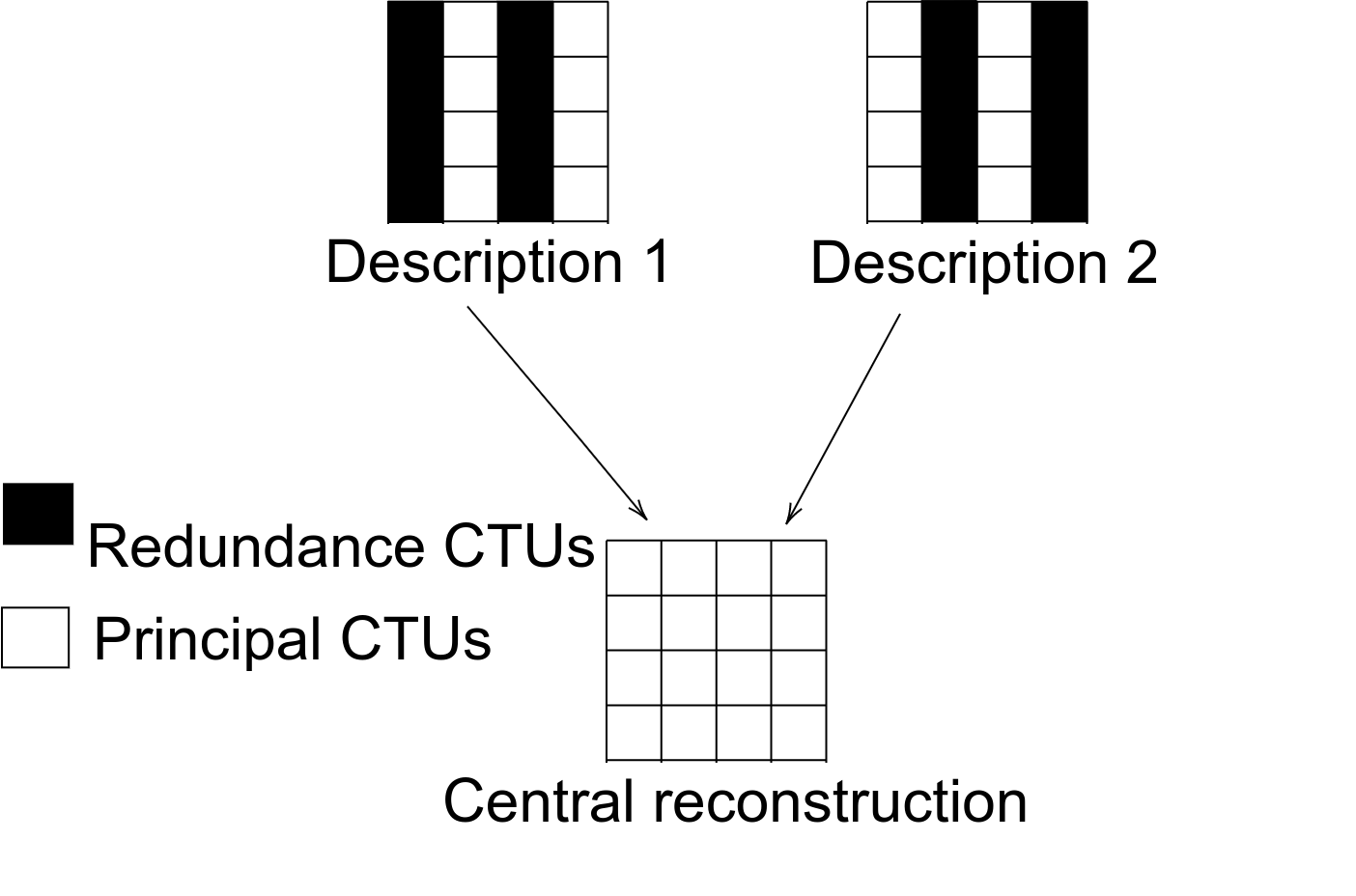}
            }
        \end{minipage}
    \caption{\footnotesize{An example of principal and redundant CTU allocation.}}
    \vspace{-1.5\baselineskip}
    \label{fig:exampleofRedundancePrincipal}
\end{figure}
\vspace{-0.5\baselineskip}
Based on the decoding workflow described in the previous section, we propose a balanced MDC with 2 descriptions. Let's assume that the packet error distribution of both channel is independently and identically distributed (i.i.d). Therefore, for each frame of the sequence, the expected distortion at the decoder is expressed as: 
\vspace{-0.5\baselineskip}
\begin{equation}
    D_{e} = (1-p_{e})\sum_{j=1}^{2}D_{p,j} + p_{e}(1-p_{e})\sum_{j=1}^{2}D_{r,j} + p_{e}^{2}D_{error}
    \label{eq:ExpectedDecoded}
\end{equation}
\noindent where $D_{p,j}$ and $D_{r,j}$ are the total distortion of the principal CTUs and the redundant CTUs respectively, in the same description $j$ of the same frame. We define the following relationship:
\vspace{-1.5\baselineskip}
\begin{equation}
    D_{p,j}+D_{r,j}=\sum_{i=1}^N d_{i,j}\ \ \forall j\in\{1,2\}
\vspace{-0.4\baselineskip}
\end{equation}

with $d_{i,j}$ the quantization distortion of a CTU$_{i,j}$, $i$ being the CTU index in a frame and $j$ the description, $N$ the total number of CTUs in each frame. $p_{e}$ is the probability of packet error. The term $D_{error}$ is the distortion when the two descriptions are lost simultaneously. In this paper, we assume that $D_{error}$ is a constant and therefore it can be omitted in the cost function.

and the problem is to find the set of $\{QP_{i,j}\}$ to use for each CTU$_{i,j}$ in a frame which minimizes the expected distortion $D_{e}$ under a frame target bit rate $R_{t}$. Thus our optimal MDC rate constrained optimization problem can be expressed as:
\vspace{-0.3\baselineskip}
\begin{mini*}|s|
{\{QP_{i,j}\}}{D_{e} \qquad \forall j\in\{1,2\} \quad \forall i\in\{1,...,N\}}
{}{}
\addConstraint{R_{j}=\dfrac{R_{t}}{2}}
\addConstraint{QP_{min} \leq QP_{i,j} \leq QP_{max}}
\end{mini*}
\vspace{-1.0\baselineskip}

\noindent where $QP_{min}$ and $QP_{max}$ are the minimum and maximum quantization parameter values defined by HEVC respectively. $R_j$ is the lateral bitrate of description $j$. $R_j$ is expressed as the sum of the bitrates $R_{i,j}$ in all the CTU$_{i,j}$ of a given description $j$ as:
\vspace{-0.7\baselineskip}
\begin{equation}
    R_j=\sum_{i=1}^{N}R_{i,j}
\end{equation}
\vspace{-0.5\baselineskip}

This problem can be solved by using standard Lagrangian approach and minimizing the following cost function:
\vspace{-0.5\baselineskip}
\begin{equation}
    J_{\lambda_{1},\lambda_{2}}(R_{1},R_{2}) = D_{e} + \sum_{j=1}^{2} \lambda_{j}(R_{j}-R_{t}/2)
\vspace{-0.5\baselineskip}
    \label{eq:costfunction}
\end{equation}

As the two descriptions are independent from each other, we can therefore establish:
\vspace{-0.5\baselineskip}
\begin{equation}
    J_{\lambda_{1},\lambda_{2}}(R_{1},R_{2}) = J_{\lambda_{1}}(R_{1}) + J_{\lambda_{2}}(R_{2})
\end{equation}
where $J_{\lambda_{j}}(R_{j})$ contains only the terms of the corresponding description $j$ and is given by:
\vspace{-0.3\baselineskip}
\begin{equation}
J_{\lambda_{j}}(R_{j})=(1-p_{e})D_{p,j} + p_{e}(1-p_{e})D_{r,j} + \lambda_{j}(R_{j}-R_{t}/2)
\end{equation}
The solution for the optimization constrained problem is given by the first order conditions, leading to:
\vspace{-0.3\baselineskip}
\begin{numcases}{}
    \dfrac{\partial J_{\lambda_{j}}(R_{j})}{\partial \lambda_{j}} = 0\label{eq:partiallam1} \qquad \forall j\in\{1,2\} \\
    \dfrac{\partial J_{\lambda_{j}(R_{j})}}{\partial R_{i,j}}=0 \qquad \forall j\in\{1,2\} \quad \forall i\in\{1,..,N\} \label{eq:partialR1}
\end{numcases}

From (\ref{eq:partiallam1}) and (\ref{eq:partialR1}), we have:
\vspace{-0.2\baselineskip}
\begin{equation}
     \dfrac{\partial J_{\lambda_{j}}(R_{j}) }{\partial \lambda_{j}} = R_{j} - \dfrac{R_{t}}{2} = 0
\end{equation}
\vspace{-0.5\baselineskip}
\begin{equation}
\dfrac{\partial J_{\lambda_{j}}(R_{j}) }{\partial R_{i,j}} =       C_{i,j}\dfrac{\partial d_{i,j}(R_{i,j})}{\partial R_{i,j}} + \lambda_{j} = 0
\label{eq:partialR1dev}
\end{equation}
where $d_{i,j}(R_{i,j})$ corresponds to the quantization distortion of the CTU$_{i,j}$ and, 
\vspace{-0.5\baselineskip}
\begin{equation}
    C_{i,j} = 
    \begin{cases}
        p_{e}(1-p_{e}) $\ for a redundant CTU$\\
        1-p_{e}  $\qquad for a principal CTU$\\
    \end{cases}
\end{equation}
The process of assigning $C_{i,j}$ in the equation (\ref{eq:partialR1dev}) is done in a complementary way between the two descriptions. Which means that if a CTU$_{i,j}$ is principal in one description $j=1$ for example, the CTU$_{i,j}$ with the same index $i$ is redundant in the other description $j=2$. This process is illustrated in the figure \ref{fig:exampleofRedundancePrincipal}. 

However, solving equation (\ref{eq:partialR1dev}) is not obvious as the rate-distortion function $d_{i,j}(R_{i,j})$ in HEVC is not continuous and then not differentiable. In \cite{DBLP:journals/spm/SullivanW98}, authors show that the relation between rate and distortion in a CTU can be approximated by an exponential function which has the following form:
\begin{equation}
d_{i,j}(R_{i,j}) = a_{i}e^{b_{i}R_{i,j}} \ \ \forall \{i,j\}
\label{eq:exponential}
\end{equation}
where the $a_{i}$ and $b_{i}$ parameters are estimated using a linear regression in each CTU before defining the two descriptions of the residual frame {coming from the prediction processes as
showed in figure \ref{fig:MyMDCSolution}}. Then, when replacing (\ref{eq:exponential}) into (\ref{eq:partialR1dev}) it comes:
\vspace{-0.5\baselineskip}

\begin{equation}
\dfrac{\partial J_{\lambda_j}(R_j) }{\partial R_{i,j}} =       C_{i,j}a_{i}b_{i}e^{b_{i}R_{i,j}} + \lambda_{j}
\end{equation}

\begin{figure*}[!htp]
\vspace{-0.5\baselineskip}
\centering
	\begin{minipage}[b]{0.3\linewidth}
		\centerline{
                \includegraphics[width = 2in]{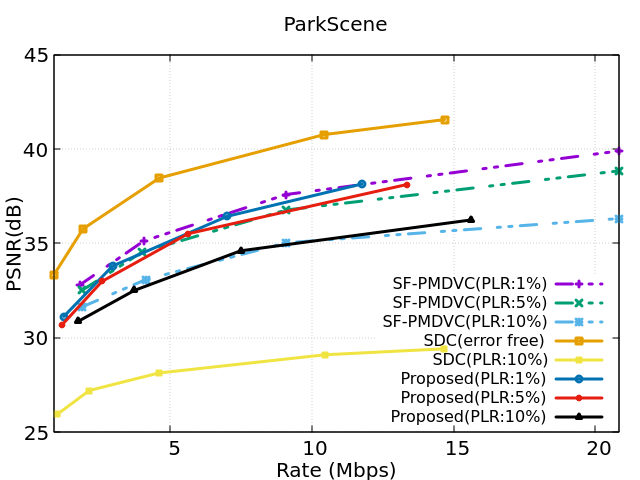}
                }
	\end{minipage}
	\begin{minipage}[b]{0.3\linewidth}
		\centerline{
            \includegraphics[width = 2in]{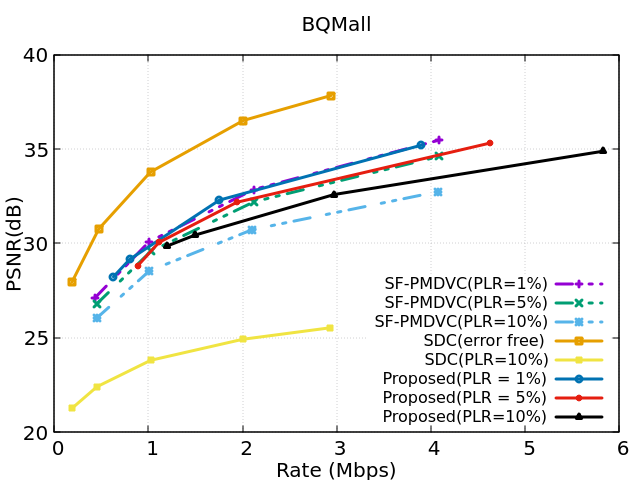}
            }
	   \end{minipage}
\centering
	\begin{minipage}[b]{0.3\linewidth}
		\centerline{
            \includegraphics[width = 2in]{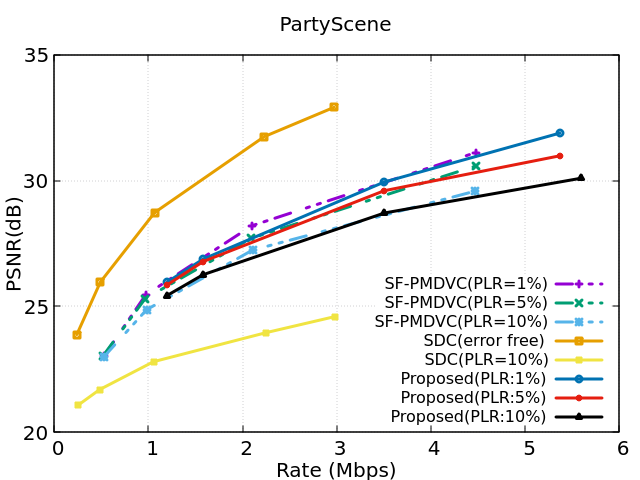}
            }
	   \end{minipage}
	\begin{minipage}[b]{0.3\linewidth}
		\centerline{
            \includegraphics[width = 2in]{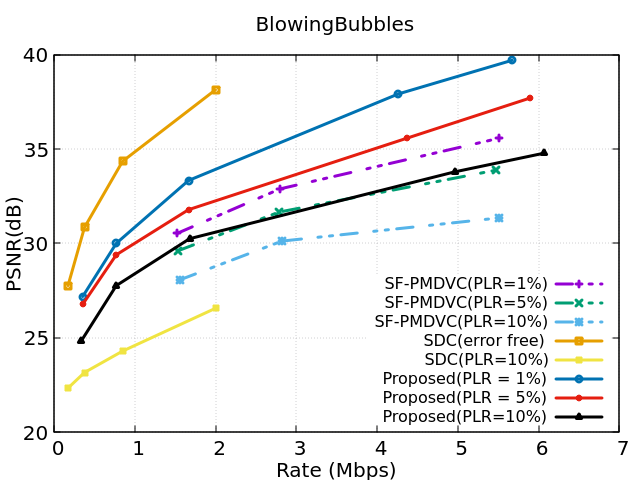}}
	   \end{minipage}
        \begin{minipage}[b]{0.3\linewidth}
            \includegraphics[width=2in]{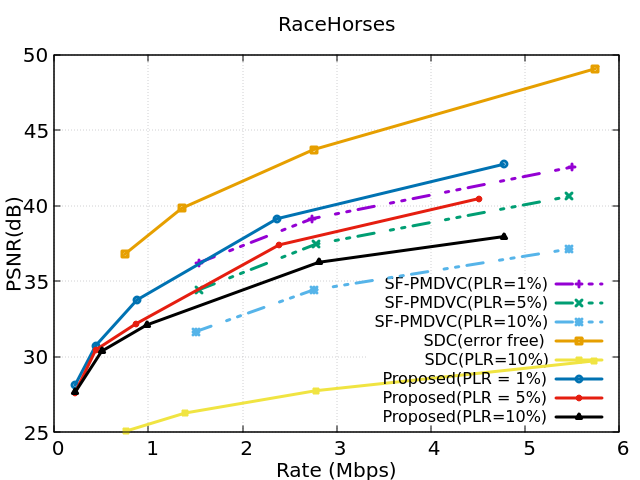}
        \end{minipage}
\vspace{-0.5\baselineskip}
\caption{\footnotesize{Comparison of the average Rate-Distortion with the method SF-PMDVC \label{fig:comparison}\cite{DBLP:journals/jvcir/WangCZC22} under packet erasure: The experiment was conducted for three different packet loss rates: 0.1, 0.05, and 0.01. Each compressed stream was simulated three times over with different packet lost rate, and the average PSNR was computed. Our proposed MDC used LD-P configuration for encoding, while the referenced method used RA. SDC stands for Single Description Coding}}
\vspace{-1.0\baselineskip}
\end{figure*}

Therefore, for each CTU$_{i,j}$, we must solve the following system of equations:
\vspace{-0.5\baselineskip}
\begin{equation}
    \begin{cases}
        R_{i_{min}}<R_{i,j} < R_{i_{max}}\\
        C_{i,j}a_{i}b_{i}e^{b_{i}R_{i,j}} + \lambda_{j} = 0\\ 
        {R_{j}} -\dfrac{R_{t}}{2}=0\\
    \end{cases}
    \forall j\in\{1,2\} \quad \forall i\in\{1,N\}
    \label{eq:systemofEquation}
\vspace{-0.3\baselineskip}
\end{equation}

where $R_{i_{min}}$ and $R_{i_{max}}$ are given by $QP_{max}$ and $QP_{min}$ repectively for each CTU$_{i,j}$.
For a fixed $\lambda_j$, we can solve the system of equation (\ref{eq:systemofEquation}) by using any bounded Newton method. In this article we used the algorithm BFGS-B \cite {DBLP:journals/toms/MoralesN11}. For each description, we need to find the value of $\lambda_{j}$ which allows to match the target bit rate $R_t$. To find the $\lambda_{j}$ value, we can use the bisection algorithm. The optimization provides the set of $\{R_{i,j}^*\}$ to be used in the CTU$_{i,j}$. The details of the optimization solution are given in algorithm \ref{algo:algoSolveLBGS}. After finding the bitrate budgets $R_{i,j}^*$ for all the CTU$_{i,j}$, we can estimate the corresponding $QP_{i,j}$ values.
\begin{algorithm}[!htp]
\caption{Solve the system of equation (\ref{eq:systemofEquation}) \label{algo:algoSolveLBGS}}
\begin{algorithmic}
\State $\text{Initialize } R_j, R_t, \lambda_{MAX}, \lambda_{MIN}, \epsilon, R_{i_{max}}, R_{i_{min}}$

\While{$|R_{j}-\frac{R_{t}}{2}| > \epsilon$}
    \State $ \lambda_{j} \gets \frac{\lambda_{MAX} + \lambda_{MIN}}{2}$
    \State $\{R_{i,j}^*\} \gets minimize (J_{\lambda_{j}}(R_{j}))$
    \State $\{R_{i,j}^*\} \gets clip(R_{i_{max}}, R_{i_{min}},R_{i,j})$
    \If {$R_{j}-\frac{R_{t}}{2} < 0$}
        \State $\lambda_{MAX} \gets \frac{\lambda_{j} + \lambda_{MIN}}{2}$
    \Else
        \State $\lambda_{MIN} \gets \frac{\lambda_{j} + \lambda_{MAX}}{2}$
    \EndIf
\EndWhile
\end{algorithmic}
\end{algorithm}
\vspace{-0.8\baselineskip}

\subsection{Adaptive IDR}
\vspace{-0.5\baselineskip}
It is very important to adapt the Instantaneous Decoder Refresh (IDR) frame in function of the packet error rate $p_{e}$. The IDR frame allows the encoder to send a intra-frame and a signal to the decoder to clear the Central Reconstructed Buffer, and all the frames can be decoded from this IDR frame. This frame prevents the error propagation between the frames during the decoding process. However, the encoder needs to select the right amount of IDR frames to achieve the best coding quality with respect to the channel noise. The study \cite{DBLP:journals/spic/CoteK99} has shown that the optimal IDR frame period under i.i.d. packet error distribution is given by: 
\vspace{-1.0\baselineskip}
\begin{equation}
    T_{IDR} = \frac{1}{p_{e}}.
\end{equation}
In our application, we use this formula to decide the right amount of IDR frames in order to have the best trade-off between quality and compression performance.
\vspace{-1.0\baselineskip}
\section{Experimental Result}
\vspace{-1.0\baselineskip}
This section presents the experimental evaluation of our proposed solution under a packet erasure channel. The solution, mentioned earlier, is implemented inside the HM codec \cite{HM}. We used the Evalvid 2.7 framework \cite{DBLP:conf/cpe/KlaueRW03} and the lost pattern defined in \cite{stephan1999error} to create packet errors in the HEVC compressed stream. The Real-Time Protocol (RTP) specified in \cite{rfc7798} was used to split the video bitstream in multiple smaller packets that are compatible with the IP network. The RTCP packet informed the encoder about the packet error rate $p_{e}$ through a feedback channel. The low-delay P (LD-P) configuration was used to encode sequences.

Figure \ref{fig:comparison} demonstrates that our proposed solution, which employs the LD-P configuration, outperforms the SF-PMDVC method with Random Access (RA) configuration for high packet error rates and high-motion sequences such as BQMall, RaceHorses, and BlowingBubbles. For other sequences, our method achieves the same performance as the referenced method. Although the RA configuration provides better compression performance than LD-P, it is impractical for real-time applications due to its complexity and high latency. On average, the RA configuration takes 40\% longer to execute than the LD-P configuration. Furthermore, the RA configuration requires at least three frames in the buffer to be encoded because of its use of B-Frames. Thus, our proposed method is better suited for real-time applications.
\vspace{-1.2\baselineskip}
\section{Conclusion}
\vspace{-1.0\baselineskip}
In this study, we proposed a spatial-based multiple description encoding bit allocation and decoding solution that is adapted to HEVC standard. Our proposed MD coding scheme includes a bit allocation that distributes the redundancy between descriptions by adjusting the QP value for each CTU within a frame based on the channel characteristics and an IDR adaptation. This solution meets the requirements of low latency and good compression performance, making it suitable for use in remote control vehicles.

In our perspective, a more robust error handling mechanism with finer grain error detection at the CU level will enhance the performance of the system. Various methods such as those discussed in \cite{le_antonini_lambert_alioua,DBLP:conf/icip/AgostiniAK09}, could be employed to improve the decoding performance. Additionally, a scheme of optimization with error mismatch propagation model should improve the performance of the system.
\vspace{-1.0\baselineskip}

\bibliographystyle{IEEEbib}

\bibliography{refs}

\begin{thebibliography}{10}

\bibitem{DBLP:journals/comsur/ParvezRGSD18}
Imtiaz Parvez, Ali Rahmati, Ismail G{\"{u}}ven{\c{c}}, Arif~I. Sarwat, and
  Huaiyu Dai,
\newblock ``A survey on low latency towards 5g: Ran, core network and caching
  solutions,''
\newblock {\em {IEEE} Commun. Surv. Tutorials}, vol. 20, no. 4, pp. 3098--3130,
  2018.

\bibitem{1698467}
C.~Soldani, G.~Leduc, F.~Verdicchio, and A.~Munteanu,
\newblock ``Multiple description coding versus transport layer fec for
  resilient video transmission,''
\newblock in {\em International Conference on Digital Telecommunications
  (ICDT'06)}, 2006, pp. 20--20.

\bibitem{DBLP:journals/tit/GamalC82a}
Abbas A.~El Gamal and Thomas~M. Cover,
\newblock ``Achievable rates for multiple descriptions,''
\newblock {\em {IEEE} Trans. Inf. Theory}, vol. 28, no. 6, pp. 851--857, 1982.

\bibitem{DBLP:journals/spm/Goyal01a}
Vivek~K. Goyal,
\newblock ``Multiple description coding: compression meets the network,''
\newblock {\em {IEEE} Signal Process. Mag.}, vol. 18, no. 5, pp. 74--93, 2001.

\bibitem{DBLP:journals/mms/KazemiSS14}
Mohammad Kazemi, Shervin Shirmohammadi, and Khosrow~Haj Sadeghi,
\newblock ``A review of multiple description coding techniques for
  error-resilient video delivery,''
\newblock {\em Multim. Syst.}, vol. 20, no. 3, pp. 283--309, 2014.

\bibitem{DBLP:journals/tcsv/TilloGO08}
Tammam Tillo, Marco Grangetto, and Gabriella Olmo,
\newblock ``Redundant slice optimal allocation for {H.264} multiple description
  coding,''
\newblock {\em {IEEE} Trans. Circuits Syst. Video Technol.}, vol. 18, no. 1,
  pp. 59--70, 2008.

\bibitem{DBLP:journals/tcsv/RadulovicFWHH10}
Ivana Radulovic, Pascal Frossard, Ye{-}Kui Wang, Miska~M. Hannuksela, and Antti
  Hallapuro,
\newblock ``Multiple description video coding with {H.264/AVC} redundant
  pictures,''
\newblock {\em {IEEE} Trans. Circuits Syst. Video Technol.}, vol. 20, no. 1,
  pp. 144--148, 2010.

\bibitem{DBLP:journals/tcsv/HsiaoT10}
Chia{-}Wei Hsiao and Wen{-}Jiin Tsai,
\newblock ``Hybrid multiple description coding based on {H.264},''
\newblock {\em {IEEE} Trans. Circuits Syst. Video Technol.}, vol. 20, no. 1,
  pp. 76--87, 2010.

\bibitem{DBLP:journals/jrtip/Psannis16}
Kostas~E. Psannis,
\newblock ``{HEVC} in wireless environments,''
\newblock {\em J. Real Time Image Process.}, vol. 12, no. 2, pp. 509--516,
  2016.

\bibitem{DBLP:series/icas/SchwarzSM14}
Heiko Schwarz, Thomas Schierl, and Detlev Marpe,
\newblock ``Block structures and parallelism features in {HEVC},''
\newblock in {\em High Efficiency Video Coding}, Integrated Circuits and
  Systems, pp. 49--90. Springer, 2014.

\bibitem{DBLP:journals/mta/MajidOA18}
Muhammad Majid, Muhammad Owais, and Syed~Muhammad Anwar,
\newblock ``Visual saliency based redundancy allocation in {HEVC} compatible
  multiple description video coding,''
\newblock {\em Multim. Tools Appl.}, vol. 77, no. 16, pp. 20955--20977, 2018.

\bibitem{DBLP:journals/jvcir/WangCZC22}
Feifeng Wang, Jing Chen, Huanqiang Zeng, and Canhui Cai,
\newblock ``Spatial-frequency {HEVC} multiple description video coding with
  adaptive perceptual redundancy allocation,''
\newblock {\em J. Vis. Commun. Image Represent.}, vol. 88, pp. 103614, 2022.

\bibitem{LABIOD201923}
Mohamed~Aymen Labiod, Mohamed Gharbi, François-Xavier Coudoux, Patrick Corlay,
  and Noureddine Doghmane,
\newblock ``Cross-layer scheme for low latency multiple description video
  streaming over vehicular ad-hoc networks (vanets),''
\newblock {\em AEU - International Journal of Electronics and Communications},
  vol. 104, pp. 23--34, 2019.

\bibitem{DBLP:conf/mmsp/SabevaJKD06}
Galina Sabeva, Salma~Ben Jamaa, Michel Kieffer, and Pierre Duhamel,
\newblock ``Robust decoding of {H.264} encoded video transmitted over wireless
  channels,''
\newblock in {\em {MMSP}}. 2006, pp. 9--13, {IEEE}.

\bibitem{DBLP:journals/spm/SullivanW98}
Gary~J. Sullivan and Thomas Wiegand,
\newblock ``Rate-distortion optimization for video compression,''
\newblock {\em {IEEE} Signal Process. Mag.}, vol. 15, no. 6, pp. 74--90, 1998.

\bibitem{DBLP:journals/toms/MoralesN11}
Jos{\'{e}}~Luis Morales and Jorge Nocedal,
\newblock ``Remark on "algorithm 778: {L-BFGS-B:} fortran subroutines for
  large-scale bound constrained optimization",''
\newblock {\em {ACM} Trans. Math. Softw.}, vol. 38, no. 1, pp. 7:1--7:4, 2011.

\bibitem{DBLP:journals/spic/CoteK99}
Guy C{\^{o}}t{\'{e}} and Faouzi Kossentini,
\newblock ``Optimal intra coding of blocks for robust video communication over
  the internet,''
\newblock {\em Signal Process. Image Commun.}, vol. 15, no. 1-2, pp. 25--34,
  1999.

\bibitem{HM}
``High {Efficiency} {Video} {Coding} ({HEVC}) {Test} {Model} 16 ({HM} 16)
  {Encoder} {Description} {Update} 10 {\textbar} {MPEG},'' .

\bibitem{DBLP:conf/cpe/KlaueRW03}
Jirka Klaue, Berthold Rathke, and Adam Wolisz,
\newblock ``Evalvid - {A} framework for video transmission and quality
  evaluation,''
\newblock in {\em Computer Performance Evaluations, Modelling Techniques and
  Tools. 13th International Conference, {TOOLS} 2003, Urbana, IL, USA,
  September 2-5, 2003, Proceedings}, Peter Kemper and William~H. Sanders, Eds.,
  2003.

\bibitem{stephan1999error}
W~Stephan,
\newblock ``Error patterns for internet video experiments,''
\newblock {\em ITU-T SG16 Document Q15-I-16-R1}, vol. 488, 1999.

\bibitem{rfc7798}
Ye-Kui Wang, Yago Sanchez, Thomas Schierl, Stephan Wenger, and Miska~M.
  Hannuksela,
\newblock ``{RTP Payload Format for High Efficiency Video Coding (HEVC)},'' RFC
  7798, Mar. 2016.

\bibitem{le_antonini_lambert_alioua}
Trung~Hieu Le, Marc Antonini, Marc Lambert, and Karima Alioua,
\newblock ``Codage vidéo à description multiple basé sur hevc pour le
  pilotage de véhicules semi-autonomes,''
\newblock in {\em GRETSI}, 2022.

\bibitem{DBLP:conf/icip/AgostiniAK09}
Marie~Andr{\'{e}}e Agostini, Marc Antonini, and Michel Kieffer,
\newblock ``Map estimation of multiple description encoded video transmitted
  over noisy channels,''
\newblock in {\em {ICIP}}. 2009, pp. 3069--3072, {IEEE}.

\end{thebibliography}
\end{document}